\documentstyle[aps,twocolumn,epsfig]{revtex}

\begin{document}
\draft

\twocolumn[\hsize\textwidth\columnwidth\hsize\csname@twocolumnfalse\endcsname

\title{Bipartite maximally entangled nonorthorgonal states}
\author{Xiaoguang Wang}
\address{Institute of Physics and Astronomy, Aarhus University, DK-8000, Aarhus C, Denmark, and}
\address{Institute for Scientific Interchange (ISI) Foundation, Viale Settimio Severo 65, 
I-10133 Torino, Italy}
\date{\today}
\maketitle
\begin{abstract}
We give conditions under which general bipartite entangled nonorthogonal states become
maximally entangled states. By the conditions we construct a large class of entangled
nonorthogonal states with exact one ebit of entanglement in both bipartite and multipartite 
systems. One remarkable property is that the amount of entanglement in this class of states is 
independent on the parameters involved in the states. Finally we discuss how to 
generate the bipartite maximally entangled nonorthogonal states.
\end{abstract}

] \narrowtext

\section{Introduction}

Quantum entanglement has generated much interest in the quantum information
processing such as quantum teleportation\cite{Tele}, superdense coding\cite
{Dense}, quantum key distribution\cite{Key}, and telecoloning\cite{Clone}.
The entangled orthogonal states receive much attention in the study of
quantum entanglement. However the entangled nonorthogonal states also play
an important role in the quantum cryptography\cite{Crypto2} and quantum
information processing\cite{Qip}. Bosonic entangled coherent states (ECS)%
\cite{Barry} and su(2) and su(1,1) ECS\cite{Wang} are typical examples of
entangled nonorthogonal states.

The coherent state can be used to encode quantum information on continuous
variables\cite{Lloyd99} and several schemes\cite{Bartlett,Jeong,Ralph} have
been proposed for realizing quantum computation. The overlap $\langle \alpha
|-\alpha \rangle $ of two coherent states $|\pm \alpha \rangle $ of $\pi $
difference is $\exp (-2|\alpha |^2),$ which decreases exponentially with $%
\alpha .$ Then we can identify the coherent states $|\pm \alpha \rangle $
with large $\alpha $ as basis states of a logical qubit:

\begin{equation}
|0\rangle _L=|\alpha \rangle ,\text{ }|1\rangle _L=|-\alpha \rangle .
\label{eq:logic}
\end{equation}
One can also use Schr\"{o}dinger cat states\cite{YS}, the even and odd
coherent states $|\alpha \rangle _{\pm }$ $=$ $\left( |\alpha \rangle \pm
|-\alpha \rangle \right) $ $/\sqrt{2(1\pm e^{-2|\alpha |^2})}$ to encode a
qubit\cite{Cochrane,Oliveira} and they are exactly orthogonal. However these
states are extremely sensitive to photon loss.

Now we consider the bosonic ECS\cite{Barry},

\begin{equation}
|\alpha ;\alpha \rangle =\frac 1{\sqrt{2(1-e^{-4|\alpha |^2})}}\left(
|\alpha \rangle \otimes |\alpha \rangle -|-\alpha \rangle \otimes |-\alpha
\rangle \right) ,  \label{eq:aa}
\end{equation}
which can be produced by using a 50/50 beam splitter. If 
\mbox{$\vert$}%
$\alpha |$ is large, the ECS is considered as a state of two logical qubits (%
\ref{eq:logic}):

\begin{equation}
|\alpha ;\alpha \rangle =\frac 1{\sqrt{2}}\left( |0\rangle _L\otimes
|0\rangle _L-|1\rangle _L\otimes |1\rangle _L\right) ,
\end{equation}
which is obviously a maximally entangled state (MES), the singlet state.

More superisingly it is found that the ECS $|\alpha ;\alpha \rangle $
possess exactly one ebit entanglement\cite{Enk} and the amount of
entanglement is independent $\alpha $. Dountlessly the ECS is a MES as it
can be rewritten as

\begin{equation}
|\alpha ;\alpha \rangle =\frac 1{\sqrt{2}}\left( |\alpha \rangle _{+}\otimes
|\alpha \rangle _{-}+|\alpha \rangle _{-}\otimes |\alpha \rangle _{+}\right)
\label{eq:alpha}
\end{equation}
in terms of the even and odd coherent states $|\alpha \rangle _{\pm }.$ Eq.(%
\ref{eq:alpha}) shows that the state $|\alpha ;\alpha \rangle $ manifestly
has one ebit of entanglement.

In this paper we give conditions under which general bipartite entangled
nonorthogonal states become MES. Using the conditions we construct a large
class of bipartite maximally entangled nonorthogonal states in both the
bipartite and multipartite systems. We also propose some methods to generate
the bipartite maximally entangled nonorthogornal states.

\section{MES condition for bipartite entangled states}

We begin with a standard general bipartite entangled state\cite{ECS,Mann}

\begin{equation}
|\Psi \rangle =\mu |\bar{\alpha}\rangle \otimes |\bar{\beta}\rangle +\nu |%
\bar{\gamma}\rangle \otimes |\bar{\delta}\rangle ,  \label{eq:psi}
\end{equation}
where $|\bar{\alpha}\rangle $ and $|\bar{\gamma}\rangle $ are {\em %
normalized states} of system 1 and similarly $|\bar{\beta}\rangle $ and $|%
\bar{\delta}\rangle $ are states of system 2 with complex $\mu $ and $\nu .$
We consider the nonorthogonal case, i.e., the overlaps $\langle \bar{\alpha}%
|\gamma \rangle $ and $\langle \bar{\beta}|\bar{\delta}\rangle $ are
nonzero. After normalization, the bipartite state $|\Psi \rangle $ is given
by

\begin{equation}
|\Psi \rangle =a|\bar{\alpha}\rangle \otimes |\bar{\beta}\rangle +d|\bar{%
\gamma}\rangle \otimes |\bar{\delta}\rangle ,
\end{equation}
where $a=\mu /N_{12},$ $d=\nu /N_{12},$ and

\begin{equation}
N_{12}=\sqrt{|\mu |^2+|\nu |^2+\mu \nu ^{*}\langle \bar{\gamma}|\bar{\alpha}%
\rangle \langle \bar{\delta}|\bar{\beta}\rangle +\mu ^{*}\nu \langle \bar{%
\alpha}|\bar{\gamma}\rangle \langle \bar{\beta}|\bar{\delta}\rangle }.
\label{eq:n12}
\end{equation}

The two nonorthogonal states $|\bar{\alpha}\rangle $ and $|\bar{\gamma}%
\rangle $ are assumed to be linearly independent and span a two-dimensional
subspace of the Hilbert space. Then we choose an orthogonal basis $\{|{\bf 0}%
\rangle ,|{\bf 1}\rangle \}$ as 
\begin{eqnarray}
|{\bf 0}\rangle &=&|\bar{\alpha}\rangle ,|{\bf 1}\rangle =(|\bar{\gamma}%
\rangle -p_1|\bar{\alpha}\rangle )/N_1\text{ for system 1,}  \nonumber \\
|{\bf 0}\rangle &=&|\bar{\delta}\rangle ,|{\bf 1}\rangle =(|\bar{\beta}%
\rangle -p_2|\bar{\delta}\rangle )/N_2\text{ for system 2,}  \label{eq:basis}
\end{eqnarray}
where

\begin{eqnarray}
p_1 &=&\langle \bar{\alpha}|\bar{\gamma}\rangle ,\text{ }N_1=\sqrt{1-|p_1|^2}%
,\text{ }  \nonumber \\
p_2 &=&\langle \bar{\delta}|\bar{\beta}\rangle ,N_2=\sqrt{1-|p_2|^2}.
\label{eq:n1}
\end{eqnarray}
Under these basis the entangled state $|\Psi \rangle $ can be rewritten as

\begin{eqnarray}
|\Psi \rangle &=&(ap_2+dp_1)|{\bf 0}\rangle \otimes |{\bf 0}\rangle 
\nonumber \\
&&+aN_2|{\bf 0}\rangle \otimes |{\bf 1}\rangle +dN_1|{\bf 1}\rangle \otimes |%
{\bf 0}\rangle ,
\end{eqnarray}
which shows that the general entangled nonorthogonal state is considered as
a state of two logical qubits.

Then it is straightforward to obtain the reduced density matrix $\rho
_{1(2)} $ and the two eigenvalues of $\rho _1$ are given by\cite{Mann}

\begin{equation}
\lambda _{\pm }=\frac 12\pm \frac 12\sqrt{1-4|adN_1N_2|^2}  \label{eq:lam}
\end{equation}
which are identical to those of $\rho _2.$ The corresponding eigenvectors of 
$\rho _{1(2)}$ is denoted by $|\pm \rangle _{1(2).}$ Then the general theory
of the Schmidt decomposition\cite{Schmidt} implies that the normalized state 
$|\Psi \rangle $ can be written as

\begin{equation}
|\Psi \rangle =c_{+}|+\rangle _1\otimes |+\rangle _2+c_{-}|-\rangle
_1\otimes |-\rangle _2  \label{eq:plusminus}
\end{equation}
with $|c_{\pm }|^2=\lambda _{\pm }.$

From Eqs.(\ref{eq:lam}) and (\ref{eq:plusminus}) we immediately know that
the condition for the state $|\Psi \rangle $ be a MES is $|2adN_1N_2|=1.$
Using Eqs.(\ref{eq:n12}) and (\ref{eq:n1}), we rewrite the condition
explicitly as ${\cal C}=1,$ where

\begin{equation}
{\cal C}=\frac{2|\mu ||\nu |\sqrt{(1-|p_1|^2)(1-|p_2|^2)}}{|\mu |^2+|\nu
|^2+\mu \nu ^{*}p_1^{*}p_2+\mu ^{*}\nu p_1p_2^{*}}.  \label{eq:cccccc}
\end{equation}

Now we show that the quantity ${\cal C}$ is exactly one measure of
entanglement, the concurrence\cite{Con} for two qubits. There are different
measures of entanglement. One simple measure is the concurrence. Since the
system 1 and 2 in the bipartite state (\ref{eq:psi}) are essentially
two-state systems, we can characterize the entanglement of bipartite state
by the concurrence. The concurrence for a pure state $|\psi \rangle \,$is
defined by ${\cal C}=|\langle \psi |\sigma _y\otimes \sigma _y|\psi
^{*}\rangle |.$ Here $\sigma _y=i(|{\bf 1}\rangle \langle {\bf 0}|-|{\bf 0}%
\rangle \langle {\bf 1}|).$ A direct calculation shows that the concurrence
of the bipartite state $|\Psi \rangle $ is just the quantity ${\cal C}$
given by Eq.(\ref{eq:cccccc}). Then the condition for the state $|\Psi
\rangle $ be a MES is that the concurrence of the state is equal to 1 as we
hoped.

For orthorgonal state, $p_1=p_2=0,$ and the concurrence ${\cal C}=2|\mu
||\nu |/(|\mu |^2+|\nu |^2)$ which obviously satisfies $0\leq {\cal C}\leq 1.
$ The state $|\Psi \rangle $ becomes a MES when $|\mu |=|\nu |=1$ as we
expected$.\,$For partly orthorgonal state, $p_1\neq 0,p_2=0,$ Eq.(\ref
{eq:cccccc}) becomes 
\begin{equation}
{\cal C}=2|\mu ||\nu |\sqrt{1-|p_1|^2}/(|\mu |^2+|\nu |^2).
\end{equation}
Then the partly orthogonal state be a MES is when the inner product $p_1=0$. 
$\,$For completely nonorthorgonal state, $p_1\neq 0$ and $p_2\neq 0.$ It is
remarkable to see that we still have possibilities to make the concurrence $%
{\cal C}$ be 1. One case for ${\cal C}=1$ is given by

\begin{eqnarray}
\mu  &=&-\nu ,\text{ }  \nonumber \\
\langle \bar{\alpha}|\bar{\gamma}\rangle  &=&\langle \bar{\delta}|\bar{\beta}%
\rangle .  \label{eq:cond}
\end{eqnarray}
The necessary and sufficient condition for the state $|\Psi \rangle $ to be
a MES is discussed in detail in another paper\cite{Fu}. We call Eq.(\ref
{eq:cond}) as the MES condition for the general state $|\Psi \rangle .$ The
MES condition (\ref{eq:cond}) immediately gives a interesting antisymmetric
MES

\begin{equation}
|\Psi _a\rangle =\frac 1{\sqrt{2(1-|\langle \bar{\alpha}|\bar{\beta}\rangle
|^2)}}\left( |\bar{\alpha}\rangle \otimes |\bar{\beta}\rangle -|\bar{\beta}%
\rangle \otimes |\bar{\alpha}\rangle \right) .  \label{eq:anti}
\end{equation}
The amount of entanglement of the state is exactly one ebit and the
entanglement is independent of the parameters involved$.$ However for a
symmetric state

\begin{equation}
|\Psi _s\rangle =\frac 1{\sqrt{2(1+|\langle \bar{\alpha}|\bar{\beta}\rangle
|^2)}}\left( |\bar{\alpha}\rangle \otimes |\bar{\beta}\rangle +|\bar{\beta}%
\rangle \otimes |\bar{\alpha}\rangle \right) ,
\end{equation}
the corresponding concurrence is ${\cal C}=\frac{1-|\langle \bar{\alpha}|%
\bar{\beta}\rangle |^2}{1+|\langle \bar{\alpha}|\bar{\beta}\rangle |^2},$
which indicates that the symmetric state is not maximally entangled except
the orthorgonal case $\langle \bar{\alpha}|\bar{\beta}\rangle =0.$ Note that
states $|\bar{\alpha}\rangle $ and $|\bar{\beta}\rangle $ are different
normalized arbitrary states. From the above discussions we see that the
relative phase plays an important role on the entanglement.

Hirota {\it et al. }\cite{Hirotaetal} have found that the state $|\Psi
_a\rangle $ is a MES. However they impose a restriction that the overlap $%
\langle \bar{\alpha}|\bar{\beta}\rangle $ is a real number. As we discussed
here, this restriction is not necessary and the states $|\bar{\alpha}\rangle 
$ and $|\bar{\beta}\rangle $ can be arbitrary. As a illustration of the
importance of complex overlap, we consider a state

\[
|\alpha ,\alpha ^{*}\rangle =\frac 1{\sqrt{2(1-|\langle \alpha |\alpha
^{*}\rangle |^2)}}\left( |\alpha \rangle \otimes |\alpha ^{*}\rangle
-|\alpha \rangle \otimes |\alpha ^{*}\rangle \right) , 
\]
which is maximally entangled. The overlap $\langle \alpha |\alpha
^{*}\rangle =e^{|\alpha |^2(e^{-i2\theta }-1)}$ is real only when $\alpha
=|\alpha |e^{i\theta }$ is real or pure imaginary. So the MES with real
overlap is a small subset of the set formed by the MES with complex overlap.

More maximailly entangled ECS can be constructed. For instance, the bosonic
coherent state $|\alpha \rangle $ can be replaced by the abstract su(2)
coherent state and su(1,1) coherent state in the state $|\alpha ;\alpha
\rangle ,$ and then obtain the corresponding su(2) and su(1,1) ECS\cite{Wang}
with one ebit of entanglement.

\section{More general entangled nonorthogonal states.}

Now we consider a more general entangled coherent state of the following type

\begin{equation}
|\Phi \rangle =a|\bar{\alpha}\rangle \otimes |\bar{\beta}\rangle +b|\bar{%
\alpha}\rangle \otimes |\bar{\delta}\rangle +c|\bar{\gamma}\rangle \otimes |%
\bar{\beta}\rangle +d|\bar{\gamma}\rangle \otimes |\bar{\delta}\rangle 
\end{equation}
where $|\alpha \rangle $ and $|\beta \rangle $ are coherent states. We
assume that the state $|\Phi \rangle $ is a normalized state. When $b=c=0,$
the state $|\Phi \rangle $ reduces to the state $|\Psi \rangle .$ One
typical useful example of this type of  states is the state generated by the
interaction

\begin{equation}
H=\chi a_1^{\dagger }a_1a_2^{\dagger }a_2,
\end{equation}
where $a_i$ and $a_i^{\dagger }$ are the annihilation and creation operators
of system $i$. After an interaction time $t=\pi /\chi ,$ from the initial
product of coherent states $|\alpha \rangle \otimes |\beta \rangle ,$ the
output state is\cite{Rice}

\begin{eqnarray}
|\phi \rangle  &=&\frac 12\left[ (|\alpha \rangle +|-\alpha \rangle )\otimes
|\beta \rangle +(|\alpha \rangle -|-\alpha \rangle )\otimes |-\beta \rangle
\right]   \nonumber \\
&=&\frac 12(|\alpha \rangle \otimes |\beta \rangle +|\alpha \rangle \otimes
|-\beta \rangle   \nonumber \\
&&+|-\alpha \rangle \otimes |\beta \rangle -|-\alpha \rangle \otimes |-\beta
\rangle )
\end{eqnarray}
This state was successfully used to construct entangled coherent--state
qubits in an ion trap\cite{Munro}.

Using the same technique of above section we can consider the state $|\Phi
\rangle $ as a two--qubit state. We write the state $\,$as the form of qubits

\begin{eqnarray}
|\Phi \rangle  &=&(ap_2+b+cp_1p_2+dp_1)|{\bf 0\rangle \otimes |0}\rangle  
\nonumber \\
&&+N_2(a+cp_1)|{\bf 0\rangle \otimes |1}\rangle   \nonumber \\
&&+N_1(d+cp_2)|{\bf 1\rangle \otimes |0}\rangle +cN_1N_2|{\bf 1\rangle
\otimes |1}\rangle .  \label{eq:phi}
\end{eqnarray}
in the basis defined in Eq.(\ref{eq:basis}).

For a general pure state

\begin{equation}
|\psi \rangle =a|{\bf 0\rangle \otimes |0}\rangle +b|{\bf 0\rangle \otimes |1%
}\rangle +c|1{\bf \rangle \otimes |0}\rangle +d|{\bf 1\rangle \otimes |1}%
\rangle ,
\end{equation}
the concurrence is given by\cite{Con}

\begin{equation}
C=2|ad-bc|.  \label{eq:ccc}
\end{equation}
From Eqs.(\ref{eq:phi}) and (\ref{eq:ccc}), the concurrence of the state $%
|\Phi \rangle $ is obtained as

\begin{equation}
C=2\sqrt{1-|p_1|^2}\sqrt{1-|p_2|^2}|ad-bc|.  \label{eq:cccc}
\end{equation}
When $b=c=0$ Eq.(\ref{eq:cccc}) reduces to Eq.(\ref{eq:cccccc}) as we
expected. From Eq.(\ref{eq:cccc}) the corresponding concurrence of the state 
$|\phi \rangle $ is simply obtained as

\begin{equation}
C=\sqrt{(1-e^{-4|\alpha |^2})(1-e^{-4|\beta |^2})}.
\end{equation}
We see that the state becomes maximally entangled state if and only if the
amplitudes $|\alpha |\gg 1$ and $|\beta |\gg 1$.$\,$

\section{Bipartite entantglement in multipartite systems}

It is more interesting to ask if we can obtain bipartite MES in multipartite
systems. A bipartite MES with even systems we can offer is

\begin{eqnarray}
|\bar{\alpha};\bar{\beta}\rangle _{2N} &=&|\bar{\alpha}\rangle \otimes
...\otimes |\bar{\alpha}\rangle \otimes |\bar{\beta}\rangle \otimes
...\otimes |\bar{\beta}\rangle -  \nonumber \\
&&|\bar{\beta}\rangle \otimes ...\otimes |\bar{\beta}\rangle \otimes |\bar{%
\alpha}\rangle \otimes ...\otimes |\bar{\alpha}\rangle  \label{eq:even3}
\end{eqnarray}
up to a normalization constant. To see the fact that this state is a MES we
consider the first $N$ systems as system 1 and the other $N$ systems as
system 2. By this observation, these two states satisfy the MES condition (%
\ref{eq:cond}), i.e., they are the MES in the sense that the concurrence $%
{\cal C}_{(12...N)(N+1,N+2...2N)}$between the first $N$ systems and the
second $N$ systems is equal to one. Of course we can construct more
complicated bipartite MES in the multipartite system according to the MES
condition.

Now we consider a ECS defined by

\begin{equation}
|\alpha ;-\alpha \rangle _N=|\alpha \rangle \otimes ...\otimes |\alpha
\rangle -|-\alpha \rangle \otimes ...\otimes |-\alpha \rangle ,
\end{equation}
For even $N,$ this state is a bipartite MES, however for odd $N,$ usually it
is not. The above state can be considered as multipartite maximally
entangled states if the coherent states $|\pm \alpha \rangle $are considered
as logical qubits as in Eq.(\ref{eq:logic}).

After normalization, the MES $|\alpha ;-\alpha \rangle _N$ is expanded as

\begin{eqnarray}
|\alpha ;-\alpha \rangle _N &=&\frac 1{\sqrt{2(1-e^{-2N|\alpha |^2})}} 
\nonumber \\
&&\times \left( |\alpha \rangle \otimes ...\otimes |\alpha \rangle -|-\alpha
\rangle \otimes ...\otimes |-\alpha \rangle \right)  \nonumber \\
&=&\frac 1{\sqrt{\sinh (N|\alpha |^2)}}  \nonumber \\
&&\sum_{n_1...n_N}^\infty \frac{\alpha ^{n_1+...+n_N}\left[
1-(-1)^{n_1+...+n_N}\right] }{2\sqrt{n_1!...n_N!}}  \nonumber \\
&&\times |n_1...n_N\rangle
\end{eqnarray}
where $|n_1...n_N\rangle =|n_1\rangle \otimes ...\otimes |n_N\rangle $ and $%
|n_k\rangle $ are Fock states of system $k$.

In the limit $|\alpha |\rightarrow 0,$ we see that only the terms with $%
n_1+...+n_N=1$ survive, and the resultant state is

\begin{eqnarray}
|\text{{\rm W}}\rangle _N=\frac 1{\sqrt{N}} &&(|100...0\rangle
+|0100...0\rangle +...  \nonumber \\
&&+|0000...1\rangle ).
\end{eqnarray}
It is interesting to see that the state is the so-called {\rm W} state \cite
{Dur1,WangpraW}. The entanglement of {\rm W} state is maximally robust under
disposal of any one of the qubits.

The state $|\alpha ;-\alpha \rangle _N$ with even $N$ is a bipartite MES and
then the {\rm W} state with even $N$ is also a bipartite MES. For instance $|%
{\rm W}\rangle _4$ can be rewritten as

\begin{equation}
|W\rangle _4=\frac 1{\sqrt{2}}(|\Psi ^{+}\rangle \otimes |00\rangle
+|00\rangle \otimes |\Psi ^{+}\rangle ),
\end{equation}
which manifestly has one ebit of entanglement. Here the state $|\Psi
^{+}\rangle $ represents one of the Bell state, i.e,

\[
|\Psi ^{+}\rangle =\frac 1{\sqrt{2}}(|01\rangle +|10\rangle ). 
\]
However the state $|\alpha ;-\alpha \rangle _N$ with odd $N$ is not a
bipartite MES. For instance, from Eq.(\ref{eq:ccc}) the concurrence between
system 1 and systems 2 and 3 of the state $|\alpha ;-\alpha \rangle _3$ is
obtained as

\begin{equation}
{\cal C}_{1(23)}=\frac{\sqrt{(1-e^{-4|\alpha |^2})(1-e^{-8|\alpha |^2})}}{%
(1-e^{-6|\alpha |^2})}.
\end{equation}
In the limit $|\alpha |\rightarrow \infty ,$ the concurrence becomes 1 as we
expected, and in the limit $|\alpha |\rightarrow 0,$ the concurrence ${\cal C%
}_{1(23)}=\frac{2\sqrt{2}}3,$ which can be understood as follows. The state $%
|\alpha ;-\alpha \rangle _3$ becomes {\rm W} state in the limit $|\alpha
|\rightarrow 0,$ and for state $|${\rm W}$\rangle _N,$ an equality\cite
{Coffman}

\begin{equation}
{\cal C}_{12}^2+{\cal C}_{13}^2+...+{\cal C}_{1N}^2={\cal C}_{1(23..n)}^2
\end{equation}
holds. For $|{\rm W}\rangle _3$ the concurrence ${\cal C}_{12}={\cal C}%
_{13}=2/3,$ therefore ${\cal C}_{1(23)}=\frac{2\sqrt{2}}3.$

In three-qubit system we can construct a bipartite MES as

\begin{eqnarray}
|\alpha ;\frac \alpha {\sqrt{2}}\rangle _3 &=&|\alpha \rangle \otimes |\frac %
\alpha {\sqrt{2}}\rangle \otimes |\frac \alpha {\sqrt{2}}\rangle  \nonumber
\\
&&-|-\alpha \rangle \otimes |-\frac \alpha {\sqrt{2}}\rangle \otimes |-\frac %
\alpha {\sqrt{2}}\rangle ,
\end{eqnarray}
In the limit $|\alpha |\rightarrow 0,$ it reduces to the MES $\frac 1{\sqrt{2%
}}(|1\rangle \otimes |00\rangle +|0\rangle \otimes |\Psi ^{+}\rangle ).$
Further in odd systems the bipartite MES is constructed as

\begin{eqnarray}
|\alpha ;\frac \alpha {\sqrt{2N}}\rangle _{2N+1} &=&|\alpha \rangle \otimes |%
\frac \alpha {\sqrt{2N}}\rangle \otimes ...\otimes |\frac \alpha {\sqrt{2N}}%
\rangle   \nonumber \\
&&-|-\alpha \rangle \otimes |\frac{-\alpha }{\sqrt{2N}}\rangle \otimes
...\otimes |\frac{-\alpha }{\sqrt{2N}}\rangle 
\end{eqnarray}
with the concurrence ${\cal C}_{1(23..2N+1)}=1,$ which results from the
identity 
\begin{equation}
\langle \alpha |-\alpha \rangle =(\langle \alpha /\sqrt{2N}|-\alpha /\sqrt{2N%
}\rangle )^{2N}
\end{equation}
and the MES condition.

\section{Generation of the entangled states}

Now we consider how to generate the bipartite maximally entangled
nonorthogonal states in both bipartite and multipartite systems.

\subsection{Bipartite entangled states}

One method is already given by Barenco {\it et al.}\cite{Barenco} and Bu\v {z%
}ek and Hillery\cite{Buzek1}, and based on controlled-SWAP gate which is
described by the following transformation

\begin{eqnarray}
|0\rangle |\bar{\alpha}\rangle |\bar{\beta}\rangle &\rightarrow &|0\rangle |%
\bar{\alpha}\rangle |\bar{\beta}\rangle ,  \nonumber \\
|1\rangle |\bar{\alpha}\rangle |\bar{\beta}\rangle &\rightarrow &|1\rangle |%
\bar{\beta}\rangle |\bar{\alpha}\rangle .
\end{eqnarray}
Let the input state of the controlled-SWAP gate is $\frac 1{\sqrt{2}}%
(|0\rangle +|1\rangle )|\bar{\alpha}\rangle |\bar{\beta}\rangle $ and we
measure the output state. If we measure the qubit on the state $|-\rangle =%
\frac 1{\sqrt{2}}(|0\rangle -|1\rangle ),$ we obtain exactly the
antisymmetric maximally entangled state $|\Psi _a\rangle $ (\ref{eq:ccc}).
The entanglement swapping method\cite{Eswap} can be used to generate
entangled coherent states in trapped-ion systems\cite{Savage90,Brune92},
which is also discussed in Ref.\cite{WangBarry}. Here we generalize the
method proposed by van Enk and Hirota\cite{Enk}, who have studied how to
generate $|\alpha ;\alpha \rangle $ by 50/50 beam splitter.

The 50/50 beam splitter is described by $B_{1,2}=e^{i\frac \pi 4%
(a_1^{\dagger }a_2+a_2^{\dagger }a_1)},$ which transforms the state $|\alpha
\rangle _1\otimes |\beta \rangle _2$ as

\begin{eqnarray}
&&B_{1,2}|\alpha \rangle _1\otimes |\beta \rangle _2  \nonumber \\
&=&|(\alpha +i\beta )/\sqrt{2}\rangle _1\otimes |(\beta +i\alpha )/\sqrt{2}%
\rangle _2.
\end{eqnarray}
Further using the phase shifter $P_2=e^{-i\frac \pi 2a_2^{\dagger }a_2}$
which makes phase shifting by $-\pi /2,$ we can have the transformation $%
{\cal B}_{1,2}=P_2B_{12}P_2,$ which transforms the coherent states as

\begin{equation}
{\cal B}_{1,2}|\alpha \rangle _1\otimes |\beta \rangle _2=|\epsilon
_{+}\rangle _1\otimes |\epsilon _{-}\rangle _2,
\end{equation}
where $\epsilon _{\pm }=(\alpha \pm \beta )/\sqrt{2}.$ Now let the input
state be $|\alpha \rangle _{1-}\otimes |\beta \rangle _2,$ i.e., the input
state is the direct product of the odd coherent state $|\alpha \rangle _{1-}$
and the coherent state $|\beta \rangle _2.$ After the transformation ${\cal B%
}_{1,2},$ we obtain the output state as

\begin{equation}
|\epsilon _{+}\rangle _1\otimes |\epsilon _{-}\rangle _2-|-\epsilon
_{-}\rangle _1\otimes |-\epsilon _{+}\rangle _2
\end{equation}
up to a normalization constant. Apply another phase shifter $e^{-i\pi
a_2^{\dagger }a_2}$ on the above state, we obtain the unnormalized state

\begin{equation}
|\epsilon _{+}\rangle _1\otimes |-\epsilon _{-}\rangle _2-|-\epsilon
_{-}\rangle _1\otimes |\epsilon _{+}\rangle _2,
\end{equation}
which is exactly of the form of $|\Psi _a\rangle $(\ref{eq:anti}). So the
two-parameter ECS is a MES independent of the two parameters $\epsilon _{\pm
}$.

From the above procedure we can see that the odd coherent state plays an
important role. If we replace the odd coherent state by the even coherent
state and repeat the procedure, the resultant state is not a MES. If we let
the input state be the product state of two odd coherent states, $|\alpha
\rangle _{1-}\otimes |\alpha \rangle _{2-},$ the resultant state is given by

\begin{equation}
|\sqrt{2}\alpha \rangle _{1+}\otimes |0\rangle _2-|0\rangle _1\otimes |\sqrt{%
2}\alpha \rangle _{2+},
\end{equation}
which is also a MES. If we replace the input state $|\alpha \rangle
_{1-}\otimes |\alpha \rangle _{2-}$ by $|\alpha \rangle _{1+}\otimes |\alpha
\rangle _{2+}$ or $|\alpha \rangle _{1-}\otimes |\alpha \rangle _{2+},$ the
resultant states are not MES.

\subsection{Multipartite entangled coherent states}

We first introduce the Kerr transformation

\begin{equation}
{\cal K}=\exp [-i\pi (a_i^{\dagger }a_i)^2/2],
\end{equation}
where $a_i$ and $a_i^{\dagger }$ are the annihilation and creation operators
of the field mode $i$, respectively. It is well know that ${\cal K}$ can
transfer a coherence state $|\alpha \rangle _i$ into a superposition of two
coherent states\cite{YS} $|\pm \alpha \rangle _i,$ i.e.,

\begin{equation}
{\cal K}_i|\alpha \rangle _i=2^{-1/2}(|\alpha \rangle _i+i|-\alpha \rangle
_i)
\end{equation}
up to a trivial global phase. This superposition state has proved to be very
useful in constructing optical analogs to Schr\"{o}dinger's cat state\cite
{YS}. Now we try to generate the following multipartite entangled coherent
state with $N$ modes

\begin{eqnarray}
|\alpha ;-\alpha \rangle _N^{^{\prime }} &=&2^{-1/2}(|\alpha \rangle
_1\otimes |\alpha \rangle _2\cdot \cdot \cdot |\alpha \rangle _N  \nonumber
\\
&&+i|-\alpha \rangle _1\otimes |-\alpha \rangle _2\cdot \cdot \cdot |-\alpha
\rangle _N).
\end{eqnarray}
This state is in fact a multipartite entangled state for continuous
variables. van Loock and Braunstein\cite{Vanloock} have used an interesting
unitary transformation ${\cal U}$ to creat multipartite entangled states of
continous variables. And later the transformation is used to realize optical
cloning machine for coherent states\cite{Braun}. The transformation ${\cal U}
$ is defined as a quantum network of beam splitters as

\begin{eqnarray}
{\cal U}_N &=&{\cal B}_{N-1,N}^{^{\prime }}\left( \sin ^{-1}\frac 1{\sqrt{2}}%
\right) {\cal B}_{N-2,N-1}^{^{\prime }}\left( \sin ^{-1}\frac 1{\sqrt{3}}%
\right)  \nonumber \\
&&\times \cdot \cdot \cdot \times {\cal B}_{1,2}^{^{\prime }}\left( \sin
^{-1}\frac 1{\sqrt{N}}\right) ,  \label{eq:U}
\end{eqnarray}
where ${\cal B}_{i-1,i}^{^{\prime }}(\theta )=\exp [\theta (a_{i-1}^{\dagger
}a_i-a_i^{\dagger }a_{i-1})]$ is also a beam splitter transformation acting
on mode $i-1$ and $i.$ The action of the beam splitter on the two modes can
be expressed as

\begin{eqnarray}
&&{\cal B}_{i-1,i}^{^{\prime }}(\theta )\left( 
\begin{array}{l}
a_{i-1}^{\dagger } \\ 
a_i^{\dagger }
\end{array}
\right) {\cal B}_{i-1,i}^{^{\prime }\dagger }(\theta )  \nonumber \\
&=&\left( 
\begin{array}{ll}
\cos \theta & -\sin \theta \\ 
\sin \theta & \cos \theta
\end{array}
\right) \left( 
\begin{array}{l}
a_{i-1}^{\dagger } \\ 
a_i^{\dagger }
\end{array}
\right) ,  \label{eq:B}
\end{eqnarray}

Directly from Eqs.(\ref{eq:U}) and (\ref{eq:B}), an useful relation is
obtained as

\begin{equation}
{\cal U}_N\sqrt{N}a_1^{\dagger }{\cal U}_N^{\dagger
}=\sum_{i=1}^Na_i^{\dagger }.  \label{eq:UU}
\end{equation}
Eq.(\ref{eq:UU}) leads to

\begin{equation}
{\cal U}_ND_1\left( \sqrt{N}\alpha \right) {\cal U}_N^{\dagger
}=\prod_{i=1}^ND_i\left( \alpha \right) ,  \label{eq:DD}
\end{equation}
where $D_i\left( \alpha \right) =\exp (\alpha a_i^{\dagger }-\alpha ^{*}a_i)$
is the displacement operator for mode $i.$

Let the intial state of the $N$--mode systems be $|\sqrt{N}\alpha \rangle
_1\otimes |0\rangle _2\cdot \cdot \cdot |0\rangle _N,$ and let the unitary
operator ${\cal U}_N{\cal K}_1$ act on it. Then we obtain

\begin{equation}
|\alpha ;-\alpha \rangle _N^{^{\prime }}={\cal U}_N{\cal K}_1|\sqrt{N}\alpha
\rangle _1\otimes |0\rangle _2\cdot \cdot \cdot |0\rangle _N.  \label{M1}
\end{equation}
That is to say, we can generate the multipartite ECS from a product $N$%
--mode state by successive application of the unitary operators ${\cal K}_1$
and ${\cal U}_N.$ The proposed scheme is simple and can be in principle
realized by the present technology.

We can also generate the multipartite {\rm W} state\cite{Dur1,WangpraW}.
From Eq.(\ref{eq:UU}) we obtain

\begin{equation}
|{\rm W}\rangle ={\cal U}_N|100...0\rangle .  \label{eq:WW}
\end{equation}
That is to say, the {\rm W} state can be directly generated by applying the
unitary operator \thinspace ${\cal U}_N$ to the state $|100...0\rangle .$

Now let us see how to produce another kind of multipartite ECS. Let the
initial state of $N$ bosonic systems be

\begin{equation}
|\Psi _0\rangle =(|\alpha \rangle _1-|-\alpha \rangle _1)\otimes |0\rangle
_2\otimes |0\rangle _3\otimes ...\otimes |0\rangle _N.
\end{equation}
By applying the transformation ${\cal B}_{N-1,N}...{\cal B}_{3,4}{\cal B}%
_{1,2}$ to the the initial state, we obtain

\begin{eqnarray}
&&{\cal B}_{N-1,N}...{\cal B}_{3,4}{\cal B}_{1,2}|\Psi _0\rangle   \nonumber
\\
&=&|\frac \alpha {2^{1/2}}\rangle _1\otimes |\frac \alpha {2^1}\rangle
_2\otimes ...\otimes |\frac \alpha {2^{i/2}}\rangle _i\otimes ...  \nonumber
\\
&&\otimes |\frac \alpha {2^{(N-2)/2}}\rangle _{N-2}\otimes |\frac \alpha {%
2^{(N-1)/2}}\rangle _{N-1}\otimes |\frac \alpha {2^{(N-1)/2}}\rangle _N 
\nonumber \\
&&-|\frac{-\alpha }{2^{1/2}}\rangle _1\otimes |\frac{-\alpha }{2^1}\rangle
_2\otimes ...\otimes |\frac{-\alpha }{2^{i/2}}\rangle _i\otimes ... 
\nonumber \\
&&\otimes |\frac{-\alpha }{2^{(N-2)/2}}\rangle _{N-2}\otimes |\frac{-\alpha 
}{2^{(N-1)/2}}\rangle _{N-1}\otimes |\frac{-\alpha }{2^{(N-1)/2}}\rangle _N.
\nonumber \\
&&
\end{eqnarray}
It is easy to check that the ${\cal C}_{1(23..N)}=1$ due to the identity 
\begin{equation}
\langle \frac \alpha {2^{1/2}}|\frac{-\alpha }{2^{1/2}}\rangle =\langle 
\frac \alpha {2^{(N-1)/2}}|\frac{-\alpha }{2^{(N-1)/2}}\rangle
\prod_{i=2}^{N-1}\langle \frac \alpha {2^{i/2}}|\frac{-\alpha }{2^{i/2}}%
\rangle .
\end{equation}
So this state is a bipartite MES. Note that here the integer $N$ can be
either even or odd.

\section{Conclusion}

In conclusion we have given conditions under which general bipartite
entangled nonorthogonal states become  MES. According to the conditions a
large class of bipartite maximally entangled nonorthogonal states are
constructed in both the bipartite and multipartite systems. A remarkable
property of these MES is that the amount of entanglement are independent of
parameters involved in the states. We also propose some methods to generate
the MES. Specifically the multipartite entangled coherent states and the
multipartite {\rm W} state are generated by quantum networks of beam
splitters.

The applications of the bipartite MES discussed in this paper are already
considered in the context of quantum teleportation of coherent states\cite
{Enk} and entangled coherent states\cite{Wangpra}. The MES are expected to
have more applications in the quantum information processings. Throughout
the paper we only consider the bipartite entanglement. The more difficult
task is to quantify the genuine multipartite entanglement\cite{Dur1,multi}
in the multipartite nonorthogonal states, which are now under consideration.

\acknowledgments
The author thanks for the helpful discussions with Klaus M\o lmer, Barry C.
Sanders, and Anders S\o rensen. This work is supported by the Information
Society Technologies Programme IST-1999-11053, EQUIP, action line 6-2-1 and
the European Project Q-ACTA.

\end{document}